\documentclass[
    ,final              ]{aipproc}

\layoutstyle{6x9}
\begin{document}

\title{Testing General Relativity With Laser Accelerated Electron Beams}



\author{L. \'A. Gergely}
{address={Department of Theoretical Physics, University of Szeged, Szeged 6720, Tisza L. krt. 84, Hungary},
altaddress={Department of Experimental Physics, University of Szeged, 6720 Szeged, D\'om t\'er 9, Hungary}}

\author{T. Harko}
{address={Department of Physics and
Center for Theoretical and Computational Physics, The University
of Hong Kong, Pok Fu Lam Road, Hong Kong, P. R. China}}

\begin{abstract}
Electron accelerations of the order of $10^{21}\;g$ obtained by laser fields
open up the possibility of experimentally testing one of the cornerstones of
general relativity, the weak equivalence principle, which states that the
local effects of a gravitational field are indistinguishable from those
sensed by a properly accelerated observer in flat space-time. We illustrate how
this can be done by solving the Einstein equations in vacuum and integrating
the geodesic equations of motion for a uniformly accelerated particle.
\end{abstract}


\maketitle

\section{Introduction}

Acceleration of electrons by intense laser pulses is of fundamental interest
in view of potential scientific and technological applications like modeling
astrophysical phenomena, realizing table-top particle accelerators, and fast
ignition in inertial confinement fusion \cite{1}. A properly injected fast
electron with charge $e$ and mass $m_{e}$ is captured by the laser field and
accelerated to high energies when the laser has an intensity parameter $%
a_{0}=eE_{0}/m_{e}\omega c\geq 100$, with $E_{0}$ and $\omega $ the
electrical field and frequency, respectively \cite{2}. At the Lawrence
Berkeley National Laboratory electrons were accelerated with an experimental
laser plasma accelerator to $1$ GeV, corresponding to an acceleration of $%
5.4\times 10^{20}\;g$, over about $3.3$ cm \cite{3}, while with a plasma
wakefield accelerator electrons achieved $8.9\times 10^{20}\;g$, gaining $42$
GeV energy over $85$ cm \cite{4}. Here $g=9.81$ m/s$^{2}$ is the
gravitational acceleration on the surface of the Earth. For comparison, the
gravitational acceleration on the surface of a neutron star with mass $%
M=3M_{\odot }$ and radius $R=10$ km is only $a\approx GM/R^{2}=4.08\times
10^{11}\;g$. Such a stellar object is quite close to the black hole limit of
general relativity.

The prospect of producing electron accelerations of the order of $10^{21}g$
by using laser fields opens up the possibility of direct experimental test
of the weak equivalence principle, which states that "The local effects of
motion in a curved space-time are indistinguishable from those of an
accelerated observer in a flat space" \cite{Dicke}. During the acceleration
process, the space-time around the electron may become curved, thus an
equivalent gravitational field may be generated. As a first approximation we
consider the motion of the accelerated particle in a general relativistic
framework. In Section 2 we solve the \textit{vacuum} Einstein gravitational
field equations, obtaining the space-time metric. The electrons move along
the geodesics of this vacuum space-time. In Section 3 these
equations of motion are integrated, and the solution is given in parametric
form. We briefly discuss our results and point out possible generalizations
in Section 4.

\section{The metric in the presence of the uniformly accelerated particle}

\label{s2}

The inertial motion of an object is contained in the variational principle $%
\delta \int dS_{i}=0$, with $dS_{i}^{2}=c^{2}dt^{2}-dx^{2}-dy^{2}-dz^{2}$
the square of the Minkowski line element, taken in the reference frame $%
x^{\alpha }=(ct,x,y,z)$. The uniformly accelerated motion is described by a
similar variational principle, $\delta \int dS_{a}=0$. As the constant
acceleration three-vector singles out a preferred direction, say the $z$%
-axis, the line element squared can be chosen as 
\begin{equation}
dS_{a}^{2}=F(z)c^{2}dt^{2}-\frac{1}{F(z)}dz^{2}-dx^{2}-dy^{2},  \label{1a}
\end{equation}%
with $F(z)$ a function of the $z$-coordinate only, to be determined by
solving Einstein's gravitational field equations.

By introducing the 4-velocity $\dot{x}^{\alpha }=dx^{\alpha }/dS_{a}$, the
variational principle can be rewritten conveniently as 
\begin{equation}
\delta \int \sqrt{g_{\alpha \beta }\dot{x}^{\alpha }\dot{x}^{\beta }}%
dS_{a}=0~.  \label{varprinc}
\end{equation}%
The Euler-Lagrange equations corresponding to this variational problem are 
\begin{equation}
\frac{d}{dS_{a}}\left( \frac{\partial }{\partial \dot{x}^{\mu }}\sqrt{%
g_{\alpha \beta }\dot{x}^{\alpha }\dot{x}^{\beta }}\right) -\frac{\partial }{%
\partial x^{\mu }}\sqrt{g_{\alpha \beta }\dot{x}^{\alpha }\dot{x}^{\beta }}%
=0~.  \label{2}
\end{equation}%
After some simple transformations Eq.~(\ref{2}) gives the geodesic equation, 
\begin{equation}
\frac{d^{2}x^{\mu }}{dS_{a}^{2}}+\Gamma _{\alpha \beta }^{\mu }\dot{x}%
^{\alpha }\dot{x}^{\beta }=0~,
\end{equation}%
where $\Gamma _{\alpha \beta }^{\mu }=\frac{1}{2}g^{\mu \sigma }\left(
g_{\alpha \sigma ,\beta }+g_{\beta \sigma ,\alpha }-g_{\alpha \beta ,\sigma
}\right) $ are the Christoffel symbols associated to the metric, and the
comma denotes the partial derivative with respect to the respective
coordinate. For the metric given by Eq.~(\ref{1a}), $g_{\mu \nu ,\lambda
}=F^{\prime }\delta _{0\mu }\delta _{0\nu }\delta _{3\lambda }+\left(
F^{\prime }/F^{2}\right) \delta _{3\mu }\delta _{3\nu }\delta _{3\lambda }$
and the Christoffel symbols become 
\begin{equation}
\Gamma _{\alpha \beta }^{\mu }=\frac{1}{2}\left[ \frac{F^{\prime }}{F}\left(
\delta ^{0\mu }\delta _{3\alpha }\delta _{3\beta }+\delta ^{0\mu }\delta
_{0\alpha }\delta _{0\beta }-\delta ^{3\mu }\delta _{3\alpha }\delta
_{3\beta }\right) +FF^{\prime }\delta ^{3\mu }\delta _{0\alpha }\delta
_{0\beta }\right] ~.  \label{2a}
\end{equation}

Next we explore Einstein's field equations in vacuum $R_{\mu \nu }=0$, where 
$R_{\mu \nu }$ is the Ricci curvature tensor\footnote{%
We follow the metric and curvature conventions of Ref. \cite{LL}.} 
\begin{equation}
R_{\mu \nu }=\Gamma _{\mu \rho ,\nu }^{\rho }-\Gamma _{\mu \nu ,\rho }^{\rho
}+\Gamma _{\mu \rho }^{\sigma }\Gamma _{\sigma \nu }^{\rho }-\Gamma _{\mu
\nu }^{\sigma }\Gamma _{\sigma \rho }^{\rho }~.
\end{equation}%
With the explicit form of the Christoffel symbols taken into account the
Ricci tensor can be written as 
\begin{equation}
R_{\mu \nu }=\left( \ln \sqrt{-g}\right) _{,\mu \nu }-\Gamma _{\mu \nu
}^{\sigma }\left( \ln \sqrt{-g}\right) _{,\sigma }-\Gamma _{\mu \nu ,\rho
}^{\rho }+\Gamma _{\mu \rho }^{\sigma }\Gamma _{\sigma \nu }^{\rho }~.
\end{equation}%
Since for the considered metric $\sqrt{-g}=1$, the differential equations
giving $F\left( x\right) $ take the simple form%
\begin{equation}
\frac{\partial \Gamma _{\mu \nu }^{\rho }}{\partial x^{\rho }}=\Gamma _{\mu
\rho }^{\sigma }\Gamma _{\sigma \nu }^{\rho }.  \label{3i}
\end{equation}%
Exploring Eq.~(\ref{2a}) we obtain 
\begin{equation}
\left( \frac{F^{\prime }}{F}\right) ^{2}=-\left( \frac{F^{\prime }}{F}%
\right) ^{\prime }~,\quad F^{\prime 2}=\left( FF^{\prime }\right) ^{\prime
}~,
\end{equation}%
which are solved for 
\begin{equation}
F\left( x\right) =C_{0}+C_{1}z,
\end{equation}%
with $C_{0}$ and $C_{1}$ arbitrary constants of integration. By requiring
that at $z=0$ the metric is manifestly Minkowskian, the metric compatible with
a uniformly accelerated particle becomes 
\begin{equation}
dS_{a}^{2}=\left( 1-\frac{2az}{c^{2}}\right) c^{2}dt^{2}-\frac{dz^{2}}{%
1-2az/c^{2}}-dx^{2}-dy^{2}~,  \label{metr}
\end{equation}%
with $a=d^{2}z/dt^{2}$ the constant, non-relativistic acceleration of the
particle (this can be seen from the classical limit of the Lagrangian). 
Despite a manifest singularity at $z=c^{2}/2a$, the metric is flat.

\section{The equation of motion of uniformly accelerated particles}

\label{3}

From Eq. (\ref{varprinc}) one can read the Lagrangian, or equivalently, its
square, leading to identical equations of motion: 
\begin{equation}
\Lambda =\left( 1-\frac{2az}{c^{2}}\right) c^{2}\dot{t}^{2}-\frac{1}{%
1-2az/c^{2}}\dot{z}^{2}\equiv 1~.  \label{La}
\end{equation}%
Here $\dot{t}=dt/dS_{a}$ and $\dot{z}=dz/dS_{a}$, respectively. We obtain 
\begin{eqnarray}
\frac{d}{dS_{a}}\left( \frac{\partial \Lambda }{\partial \dot{z}}\right) 
&=&-\frac{2}{1-2az/c^{2}}\ddot{z}-\frac{2a}{c^{2}}\frac{1}{\left(
1-2az/c^{2}\right) ^{2}}\dot{z}^{2}~, \\
\frac{\partial \Lambda }{\partial z} &=&-\frac{2a}{c^{2}}\left[ \frac{1}{%
1-2az/c^{2}}+\frac{2}{\left( 1-2az/c^{2}\right) ^{2}}\dot{z}^{2}\right] ~.
\end{eqnarray}%
In the second equation we have eliminated $\dot{t}^{2}$ in favor of $\dot{z}%
^{2}$, by employing the second equality in (\ref{La}). The Euler-Lagrange equation simplifies to%
\begin{equation}
\frac{d^{2}z}{dS_{a}^{2}}=\frac{a}{c^{2}}~,
\end{equation}%
showing that $z$ is quadratic in $S_{a}$. By choosing the initial conditions $%
z(0)=0$ and $\left( dz/dS_{a}\right) |_{(t=0,z=0)}=0$, we obtain the
parametric solution 
\begin{equation}
z(t)=\frac{c^{2}}{2a}\left( \frac{e^{2at/c}-1}{e^{2at/c}+1}\right)
^{2},\quad S_{a}(t)=\frac{c^{2}}{a}\frac{e^{2at/c}-1}{e^{2at/c}+1}~.
\end{equation}%
At large accelerations $e^{2at/c}>>1,~z\rightarrow c^{2}/2a$ and $%
dz/dS_{a}=\left( e^{2at/c}-1\right) /\left( e^{2at/c}+1\right) \rightarrow 1$%
. This is when the metric becomes singular. For a particle moving at a
constant acceleration of $a=2\times 10^{20}$ cm/s$^{2}$ (achievable by laser
fields \cite{3,4}), this happens quite fast, at $z=4.5$ cm.

\section{Discussions and final remarks}

\label{4}

Laser physics experiments generating large accelerations could in principle
produce intense equivalent gravitational fields, even black holes. In this
paper we have explored the metric and free motion of a test particle in
vacuum. The uniformly accelerated particle then leads to a flat space-time,
with a manifest singularity at $z=c^{2}/2a$. A more realistic description of
the acceleration of electron by laser beams could be achieved by including
the accelerating electromagnetic field with energy-momentum tensor $T_{\mu
\nu }$ as a source of the Einstein equation, $R_{\mu \nu }=\left( 8\pi
G/c^{2}\right) T_{\mu \nu }$, in this case the space-time will not be flat; also modifying correspondingly the equation of
motion, $m_{0}c(du^{\mu }/ds+\Gamma _{\alpha \beta }^{\mu }u^{\alpha
}u^{\beta })=(e/c)u_{\nu }F^{\mu \nu }$. This case deserves future
investigation.

\section{Acknowledgments}

L\'{A}G was supported by the Project "T\'AMOP-4.2.1/B-09/1/KONV-2010-0005 - Creating the Center of Excellence at the University of Szeged" of the European Union, and co-financed by the European Regional Development Fund, also by the Hungarian Scientific Research Fund (OTKA) grant no. 81364.

\end{document}